\documentclass[english]{aa}
\usepackage{mathptmx}
\usepackage[T1]{fontenc}
\usepackage[latin9]{inputenc}
\usepackage{url}
\usepackage{graphicx}
\usepackage{amssymb}
\usepackage[authoryear]{natbib}

\newcommand{\noun}[1]{\textsc{#1}}
\providecommand{\tabularnewline}{\\}

\usepackage{babel}

\begin{document}
\abstract{}{We study the spectral classification of emission-line
galaxies as star-forming galaxies or Active Galactic Nuclei (AGNs).
From the Sloan Digital Sky Survey (SDSS) high quality data, we define
an improved classification to be used for high redshift galaxies}{We
classify emission-line galaxies of the SDSS according to the latest
standard recipe using {[}O\noun{iii}{]}$\lambda$5007, {[}N\noun{ii}{]}$\lambda$6584,
{[}S\noun{ii}{]}$\lambda$6717+6731, H$\alpha$, and H$\beta$ emission
lines. We obtain four classes: star-forming galaxies, Seyfert 2, LINERs,
and composites. We then examine where these galaxies fall in the blue
diagram used at high redshift (i.e. log({[}O\noun{iii}{]}$\lambda$5007/H$\beta$)
vs. log({[}\noun{Oii}{]}$\lambda\lambda$3726+3729/H$\beta$).}{We
define new improved boundaries in the blue diagram for star-forming
galaxies, Seyfert 2, LINERs, SF/Sy2, and SF-LIN/comp classes. We maximize
the success rate to 99.7\% for the detection of star-forming galaxies,
to 86\% for the Seyfert 2 (including the SF/Sy2 region), and to 91\%
for the LINERs. We also minimize the contamination to 16\% in the
region of star-forming galaxies. We cannot reliably separate composites
from star-forming galaxies and LINERs, but we define a SF/LIN/comp
region where most of them fall (64\%).}{}


\title{Spectral classification of emission-line galaxies from the Sloan
Digital Sky Survey}

\subtitle{I. An improved classification for high redshift galaxies}

\author{F. Lamareille}

\institute{Laboratoire d\textquoteright{}Astrophysique de Toulouse-Tarbes, Université
de Toulouse, CNRS, 14 avenue Edouard Belin, F-31400 Toulouse, France\\
\email{flamare@ast.obs-mip.fr}}

\date{Received ; Accepted}

\maketitle

\section{Introduction}

Spectral classification of emission-line galaxies at low redshift
is now routinely done\textbf{ }with high quality calibrations. Using
a set of five strong emission lines -- {[}O\noun{iii}{]}$\lambda$5007,
{[}N\noun{ii}{]}$\lambda$6584, {[}S\noun{ii}{]}$\lambda$6717+6731,
H$\alpha$, and H$\beta$ --, one can reliably distinguish star-forming
galaxies, Seyfert 2 galaxies, Low Ionization Nuclear Emission Region
\citep[hereafter LINER, see][]{1980A&A....87..152H}, and composite
galaxies with both star-forming regions and an active galactic nucleus
(hereafter AGN). The equations to do such classification have been
derived successively by several authors \citep[among others]{1981PASP...93....5B,1987ApJS...63..295V,2001ApJ...556..121K,2003MNRAS.346.1055K,2006MNRAS.372..961K}.

At redshifts greater than $z\approx0.4$, the {[}N\noun{ii}{]}$\lambda$6584,
{[}S\noun{ii}{]}$\lambda$6717+6731, and H$\alpha$ emission lines
get redshifted out of the wavelength range of all major optical spectroscopic
surveys. Therefore, diagnostic diagrams need to based only on emission
lines observed in the blue part of the spectra: {[}O\noun{iii}{]}$\lambda$5007,
{[}\noun{Oii}{]}$\lambda\lambda$3726+3729, and H$\beta$. Such diagrams,
which have been used in the past e.g. by \citet{1996MNRAS.281..847T}
or \citet{1997MNRAS.289..419R}, and which we call the {}``blue diagram'',
have been recently studied again by \citet{2004MNRAS.350..396L}.\textbf{
}The latter have derived, from the 2dFGRS data, equations to distinguish
star-forming galaxies from AGNs. They have also shown that a region
exists in this diagram where both star-forming galaxies and AGNs fall
(hereafter the {}``uncertainty region'') and thus cannot be distinguished.

With the high quality Sloan Digital Sky Survey Data Release 7 (hereafter
SDSS DR7) data, it is now possible to revisit the blue diagram, and
to derive new equations more compatible with the latest red classification
by \citet{2006MNRAS.372..961K} than the ones given in \citet{2004MNRAS.350..396L}.
We derive in particular in this paper more precise boundaries between
the star-forming and AGN regions, new boundaries for the regions where
AGNs or composites are mixed with star-forming galaxies, and new equations
to distinguish between Seyfert 2 galaxies and LINERs in the blue diagram.

All spectral classifications, associated numbers, and figures presented
in this paper have been done with the \emph{JClassif} spectral classification
pipeline, which is freely available at the following website: \url{http://www.ast.obs-mip.fr/users/flamare/galaxie/}.
This paper is organized as follow: we recall the current classification
scheme in Sect.~\ref{sec:The-current-classification}, and apply
it to SDSS DR7 data. Then, we define our new improved classification
for high redshift galaxies in Sect.~\ref{sec:The-improved-classification}.
Finally we show an example application our of our new classification
in Sect.~\ref{sec:An-example-application}.

\section{The current classification scheme\label{sec:The-current-classification}}

\subsection{Data selection and the red classification}

\begin{figure}
\begin{centering}
\includegraphics[width=0.49\textwidth]{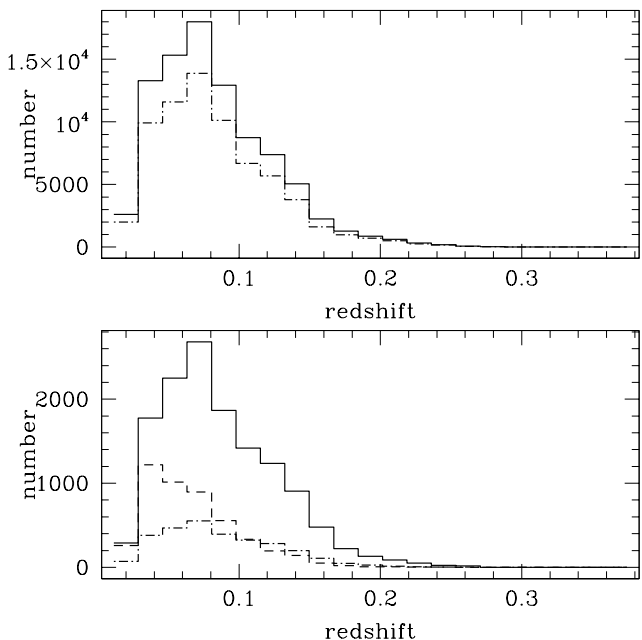}
\par\end{centering}

\caption{Redshift histograms of our data sample. \emph{Top}: The solid line
is for the whole sample, the dashed-dotted line is for star-forming
galaxies. \emph{Bottom}: The solid line is for composites, the dashed
line for LINERs, and the dashed-dotted line for Seyfert 2.}

\label{fig:histoz}
\end{figure}

\begin{figure*}
\begin{centering}
\includegraphics[width=0.49\textwidth]{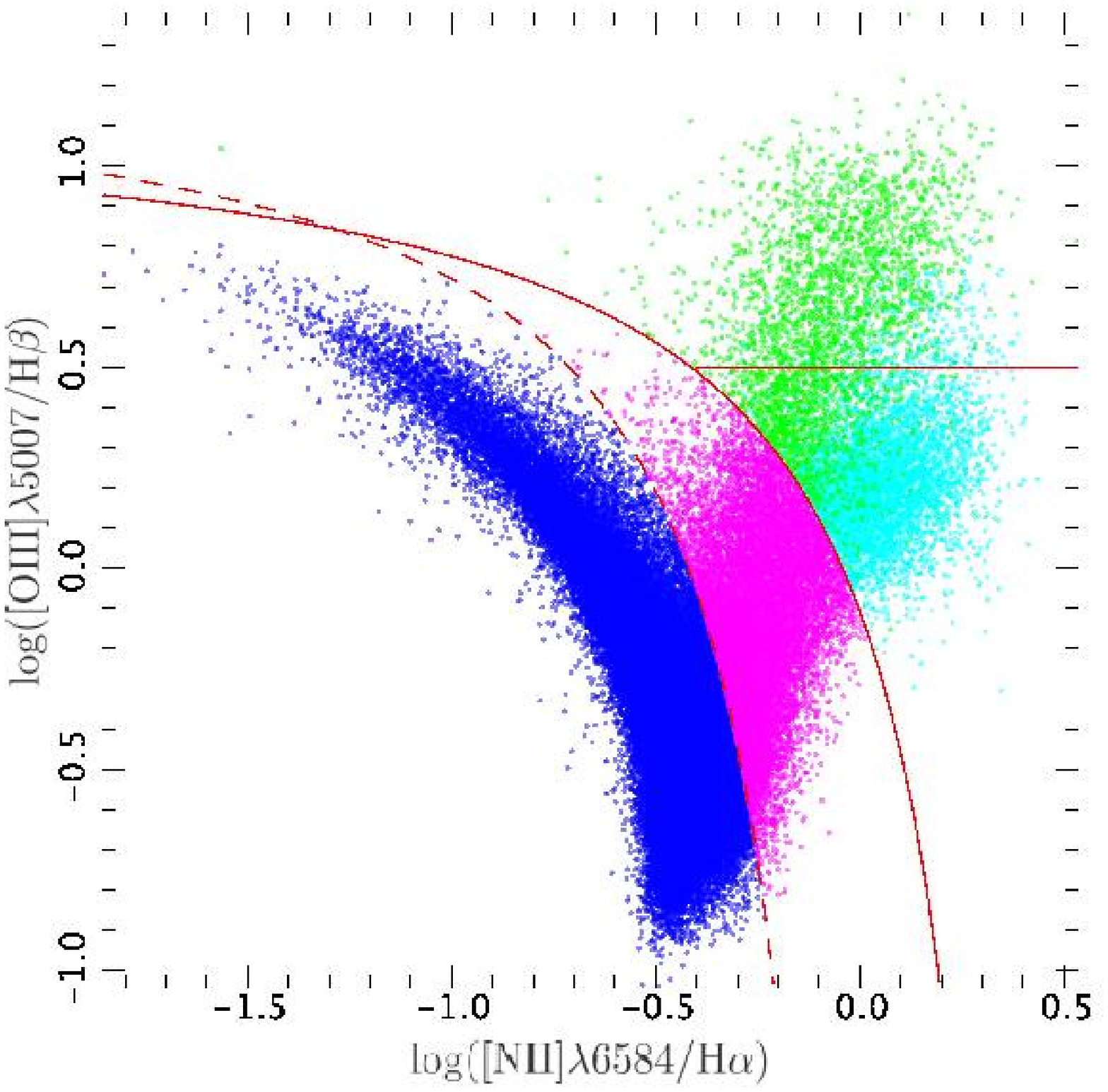} \includegraphics[width=0.49\textwidth]{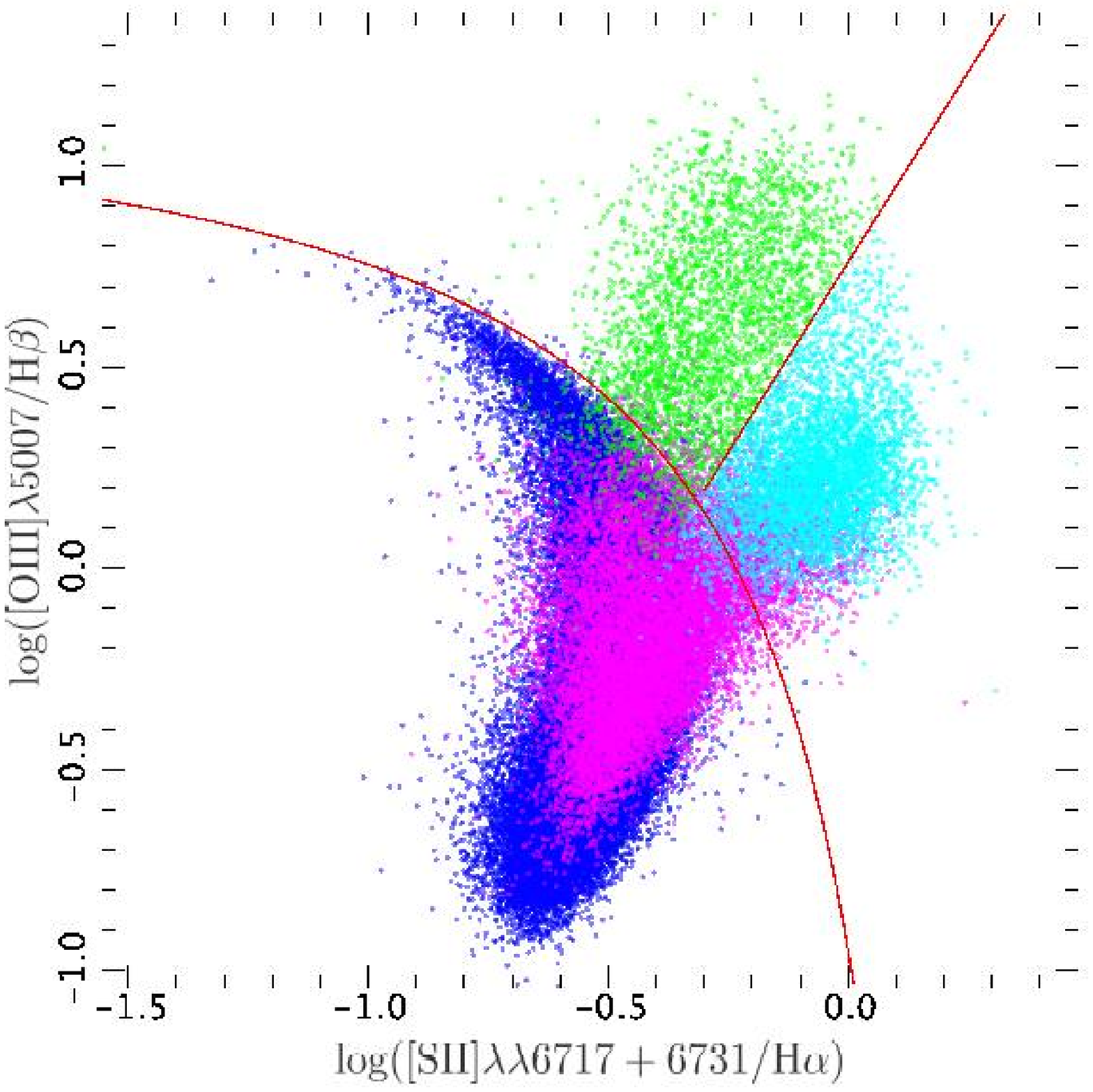} 
\par\end{centering}

\caption{This is the reference {}``red'' classification of emission-line
galaxies at low redshift. The two diagnostic diagrams show the relation
between two line ratios: log({[}O\noun{iii}{]}$\lambda$5007/H$\beta$)
vs. log({[}\noun{Nii}{]}$\lambda$6584/H$\alpha$) (left) and log({[}O\noun{iii}{]}$\lambda$5007/H$\beta$)
vs. log({[}\noun{Sii}{]}$\lambda\lambda$6717+6731/H$\alpha$) (right).
Star-forming galaxies are shown in blue, composites in magenta, Seyfert
2 in green, and LINERs in cyan. The red curves show the empirical
or theoretical separations: the solid curve (left and right) is \citet{2001ApJ...556..121K},
the dotted curve (left) is \citet{2003MNRAS.346.1055K}, the horizontal
line (left) is \citet{1987ApJS...63..295V}, and the solid line (right)
is \citet{2006MNRAS.372..961K}.}

\label{fig:refclassif}
\end{figure*}

We use SDSS DR7 emission-line measurements of 868\,492 galaxies in
the redshift range $0.0<z\lesssim0.2$. These data are available online
at the following address: \url{http://www.mpa-garching.mpg.de/SDSS/DR7/}.
The measurements are available for 927\,552 different spectra, of
which 109\,219 spectra are duplicated (twice or more) observations
of the same galaxy. We have averaged the measurements of duplicated
spectra in order to increase the signal-to-noise ratio. Measurements
which do not increase the averaged signal-to-noise ratio have been
discarded. We select emission-line galaxies with the following criterion:
the signal-to-noise ratio in the equivalent width of the emission
lines used in our study must be greater than 5. The necessary emission
lines to derive a spectral classification at low redshift are {[}O\noun{iii}{]}$\lambda$5007,
{[}N\noun{ii}{]}$\lambda$6584, {[}S\noun{ii}{]}$\lambda$6717+6731,
H$\alpha$, and H$\beta$. We also need the {[}\noun{Oii}{]}$\lambda\lambda$3726+3729
emission line which will be used to derive our new high redshift classification.
We end up with 89\,379 emission-line galaxies with the desired minimum
signal-to-noise ratio. We sort these galaxies into four classes according
to \citet{2006MNRAS.372..961K} classification scheme. We end up with
the following numbers: 67\,778 star-forming galaxies, 2\,949 Seyfert
2, 4\,912 LINERs, and 13\,740 composites. Figure~\ref{fig:histoz}
shows the redshift histograms of the four classes of emission-line
galaxies. We find that the targeted population has the same dependence
on redshift for each of the four classes, with a peak around $z\sim0.07$,
except for the LINERs, whose proportion increases at low redshift
as compared to the other classes. This possible bias has to be noted,
even if it does not affect the classification derived in this paper
that is not primarily based on relative proportions between classes.

Figure~\ref{fig:refclassif} shows this classification in the standard
BPT diagrams. In the left diagram, we remark the difference between
the old classification of Seyfert 2 and LINERs \citep{1987ApJS...63..295V}
and the new one defined by \citet{2006MNRAS.372..961K}. The right
diagram cannot be used to distinguish star-forming galaxies from composites,
which fall in the same region of this diagram. This effect has been
clearly explained by \citet{2006MNRAS.371..972S} with photoionization
models: the {[}O\noun{iii}{]}$\lambda$5007/H$\beta$ vs. {[}\noun{Nii}{]}$\lambda$6584/H$\alpha$
diagnostic diagram is the only one where composites and LINERs clearly
separate from star-forming galaxies thanks to the so-called {}``seagull
wings''. The same applies to LINERs and composites which are not
clearly separated in the {[}O\noun{iii}{]}$\lambda$5007/H$\beta$
vs. {[}S\noun{ii}{]}$\lambda$6717+6731/H$\alpha$ diagnostic diagram.
Unlike \citet{2006MNRAS.372..961K}, we did not classify as {}``ambiguous''
the composites which fall in the LINERs region, since this latter
diagram is not accurate in separating composites and LINERs.

\subsection{The blue classification}

\begin{figure}
\begin{centering}
\includegraphics[width=0.49\textwidth]{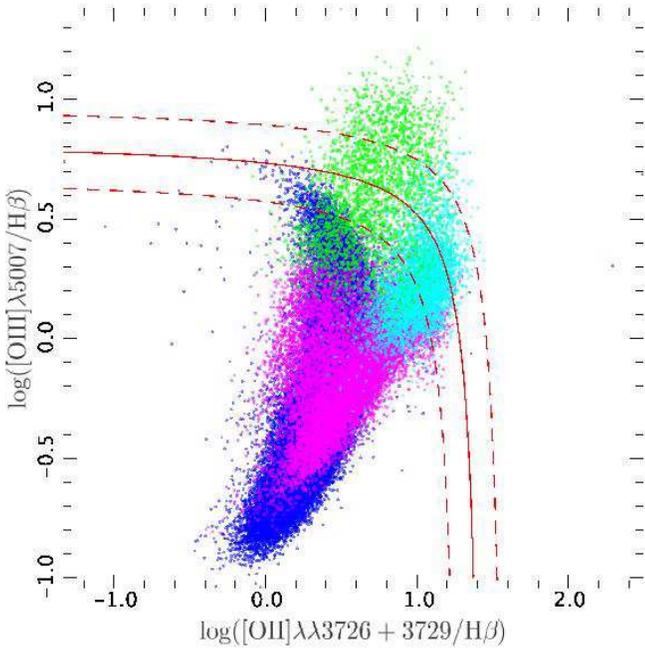}
\par\end{centering}

\caption{The {}``blue'' classification of emission-line galaxies at high
redshift. The diagnostic diagram show the relation between two line
ratios: log({[}O\noun{iii}{]}$\lambda$5007/H$\beta$) vs. log({[}\noun{Oii}{]}$\lambda\lambda$3726+3729/H$\beta$).
According to the red classification (see Fig.~\ref{fig:refclassif}),
star-forming galaxies are shown in blue, composites in magenta, Seyfert
2 in green, and LINERs in cyan. The red curves show the empirical
separations defined by \citet{2004MNRAS.350..396L}: the solid curve
is the separation between star-forming galaxies and AGNs, the dashed
curves show the uncertainty region.}

\label{fig:blueclassif}
\end{figure}

We now derive the blue classification and sort the emission-line galaxies
into one of the four classes defined by \citet{2004MNRAS.350..396L}
with the following result : 83\,654 secure star-forming galaxies,
699 secure AGNs, 3\,670 candidate star-forming galaxies, and 1\,356
candidate AGNs.

Figure~\ref{fig:blueclassif} shows where the galaxies classified
with the red classification fall in the blue diagram. The weaknesses
of the blue classification as compared to the \citet{2006MNRAS.372..961K}
classification scheme applied on the same data are evident in Fig.~\ref{fig:blueclassif},
thanks to the high quality of SDSS's line measurement software, and
to our signal-to-noise cut.\textbf{ }The blue classification is clearly
strongly biased against LINERs, which are classified for the majority
of them as star-forming galaxies or candidate star-forming galaxies
in the blue diagram. The uncertainty region is actually dominated
by AGNs (83\%), while a non-negligible number of Seyfert 2 and LINERs
(38\%) are misclassified as {}``secure'' star-forming galaxies.
Composites were classified as star-forming galaxies by \citet{2004MNRAS.350..396L}
on 2dFRGS data. This has lead them to choose an empirical separation
that goes more to the right of the blue diagram than necessary.

Nevertheless, the empirical separation defined by \citet{2004MNRAS.350..396L}
from 2dFGRS data (the solid curve) does not follow the actual boundary
between star-forming galaxies and AGNs, as seen with SDSS DR7 data.
This separation may then be improved. We may also define a new uncertainty
region, and a separation between Seyfert 2 and LINERs. The composites
cannot be distinguished from star-forming galaxies, since they fall
in the same region of the blue diagram. As mentioned above, this trend
is also present in the log({[}O\noun{iii}{]}$\lambda$5007/H$\beta$)
vs. log({[}S\noun{ii}{]}$\lambda\lambda$6717+6731/H$\alpha$) diagnostic
diagram (see Fig.~\ref{fig:refclassif} right). Unfortunately, it
cannot be avoided at high redshift, without the {[}N\noun{ii}{]}$\lambda$6584
emission-line measurement.

The red classification is not sensitive to reddening since it uses
ratios of emission lines which are close in wavelength. Conversely,
the blue classification uses a line ratio -- {[}\noun{Oii}{]}$\lambda\lambda$3726+3729/H$\beta$
-- involving two lines which are not close in wavelength. Using equivalent
widths instead of fluxes, as in this paper, removes direct dependence
on reddening. Still, the reddening does not impact exactly in the
same way the flux of emission lines and the flux of the underlying
stellar continuum. There is thus an indirect dependence of the {[}\noun{Oii}{]}$\lambda\lambda$3726+3729/H$\beta$
line ratio with reddening when calculated with equivalent widths.
Anyway this dependence is greatly reduced as compared to ratios of
line fluxes.

\section{The improved classification\label{sec:The-improved-classification}}

We now define the new improved blue classification of emission-line
galaxies. Figure~\ref{fig:newclassif} shows how the objects of different
classes, according to the red classification, fall in the new blue
diagram.

\subsection{The new star-forming -- AGN boundary}

We define a new boundary that follows more precisely the star-forming
galaxies region, as did \citet{2003MNRAS.346.1055K} compared to the
\citet{2001ApJ...556..121K} boundary in the log({[}O\noun{iii}{]}$\lambda$5007/H$\beta$)
vs. log({[}\noun{Nii}{]}$\lambda$6584/H$\alpha$) diagnostic diagram.
According to the old blue classification, 87\,324 galaxies were classified
as secure or candidate star-forming galaxies. But 19\% of them are
actually not star-forming galaxies according to the red classification.
We can reduce this contamination by excluding almost all LINERs and
most of the Seyfert 2 with a more conservative separation.

The equation that minimizes the contamination is :\begin{equation}
\log([\mathrm{O}\mathsc{iii}]/\mathrm{H}\beta)=\frac{0.11}{\log([\mathrm{O}\mathsc{ii}]/\mathrm{H}\beta)-0.92}+0.85.\label{eq:new1}\end{equation}
It corresponds to the solid curve in Fig.~\ref{fig:newclassif}.
Star-forming galaxies are below this curve, AGNs are above. The contamination
is minimized to 16\%. The minimization was done by eye, keeping in
mind to maximize at the same time the success rate. Figure~\ref{fig:blueclassif}
that a majority of composites and a number of Seyfert 2 cannot be
excluded from the region of star-forming galaxies, which explains
why the contamination\textbf{ }cannot be zero. We check that 99.7\%
of the star-forming galaxies, according to the red classification,
are correctly classified with the new blue classification, which is
quite satisfactory.

\subsection{\textbf{The mixed regions}}

Even if almost all star-forming galaxies can be correctly classified
using the blue diagram, we know that all star-forming galaxies classified
according to the blue classification are not actual star-forming galaxies.
As shown in the right panel of Fig.~\ref{fig:newclassif}, a non-negligible
number of Seyfert 2 galaxies fall into the region of star-forming
galaxies. From this plot, we easily define the boundary of the region
where star-forming galaxies become mixed with Seyfert 2:\begin{equation}
\log([\mathrm{O}\mathsc{iii}]/\mathrm{H}\beta)>0.3.\label{eq:new2}\end{equation}
This is the horizontal line in Fig.~\ref{fig:newclassif}. We call
SF/Sy2 all the galaxies above this line. Counting the region of AGNs
(as defined by Eq.~\ref{eq:new1}) and the region of SF/Sy2, 86\%
of actual AGNs are correctly classified with our new blue classification
(59\% as Seyfert 2, 26\% as SF/Sy2). The region of SF/Sy2 is nevertheless
dominated by star-forming galaxies (74\%). The left panel of Fig.~\ref{fig:newclassif}
shows that, unlike Seyfert 2, LINERs do not significantly get mixed
with star-forming galaxies: 91\% of them are correctly classified
without the need to define a SF/LINER region.

We need also to consider the case of composites which fall in the
region of star-forming galaxies and LINERs in the blue diagram: 85\%
of the composites are classified as star-forming galaxies, and 16\%
as LINERs in our new classification. We see from the right panel of
Fig.~\ref{fig:newclassif} that almost all the composites fall by
chance below the line defined in Eq.~\ref{eq:new2}. Thus this line
can be used to define the region where one should expect to find composites.
However the majority of the composites fall in a much narrower region.
We define this region with the two following inequalities:\begin{equation}
\left\{ \begin{array}{lll}
y & \le & -(x-1.0)^{2}-0.1x+0.25\\
y & \ge & (x-0.2)^{2}-0.6\end{array}\right.\label{eq:ineq}\end{equation}
where $y=\log([\mathrm{O}\mathsc{iii}]/\mathrm{H}\beta)$, and $x=\log([\mathrm{O}\mathsc{ii}]/\mathrm{H}\beta)$.
We call SF-LIN/comp all the galaxies in this region, which straddles
over the star-forming galaxies and the LINERs. 64\% of the actual
composites fall in our SF-LIN/comp region. The SF-LIN/comp region
is composed by 79\% of star-forming galaxies, 19\% composites, and
2\% LINERs.

\subsection{The new Seyfert 2 -- LINER boundary}

\begin{figure*}
\begin{centering}
\includegraphics[width=0.49\textwidth]{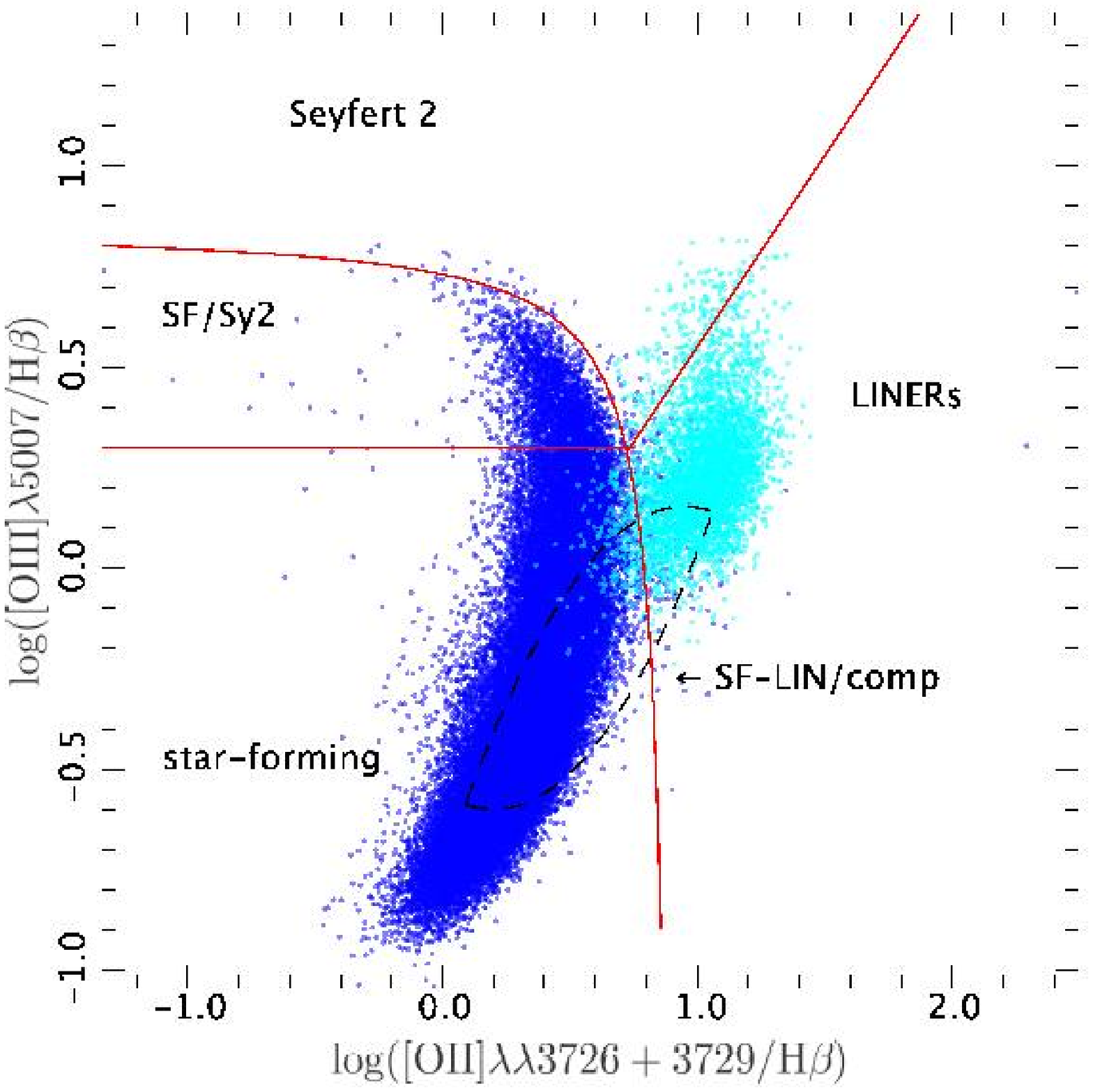} \includegraphics[width=0.49\textwidth]{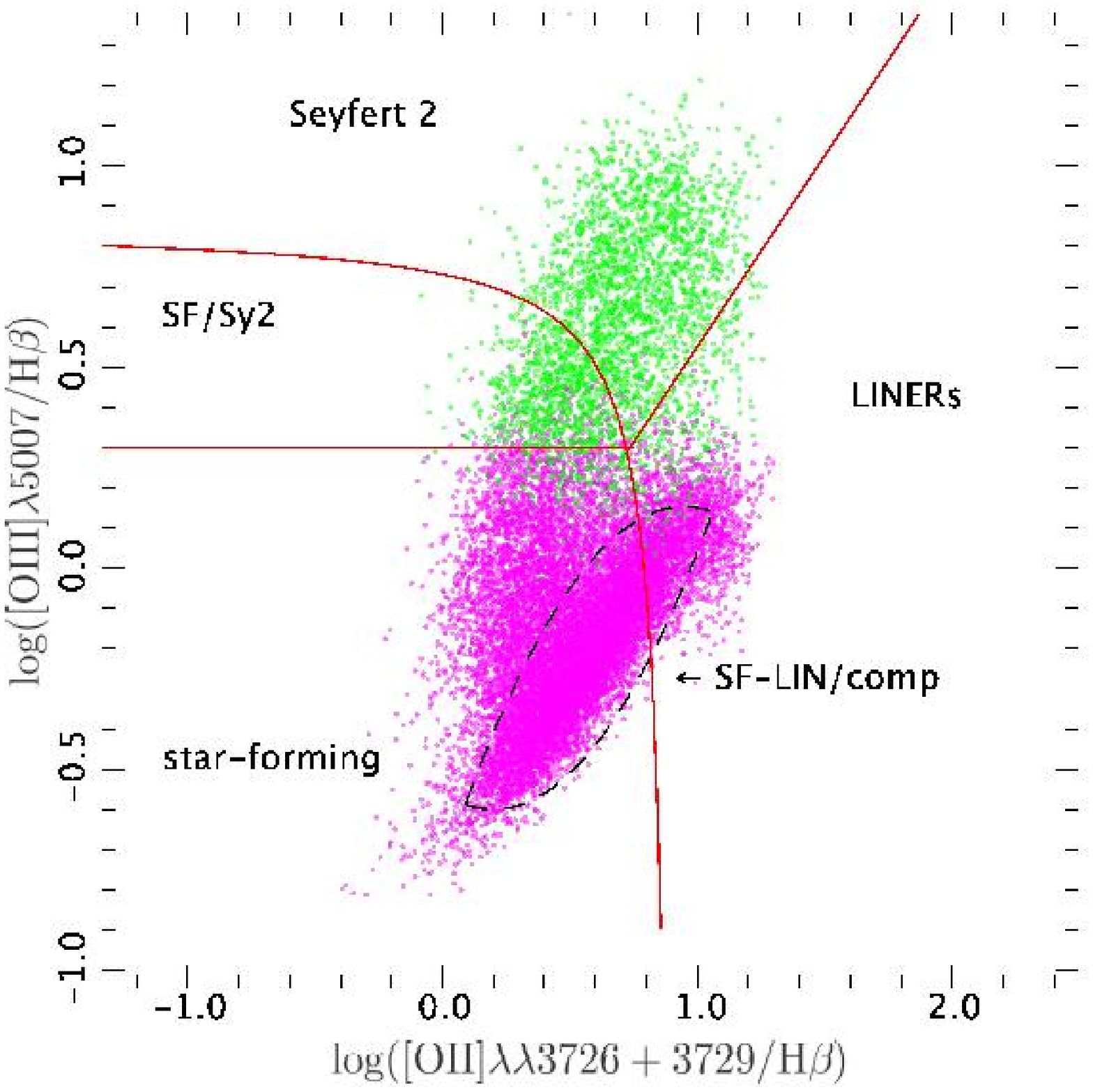} 
\par\end{centering}

\caption{This is the new improved {}``blue'' classification of emission-line
galaxies. The two diagnostic diagrams show the relation between two
line ratios: log({[}O\noun{iii}{]}$\lambda$5007/H$\beta$) vs. log({[}\noun{Oii}{]}$\lambda\lambda$3726+3729/H$\beta$).\emph{
}According to the red classification (see Fig.~\ref{fig:refclassif}),
star-forming galaxies are shown in blue, LINERs in cyan, composites
in magenta, and Seyfert 2 in green. For clarity, the two first classes
are shown only in the left panel, while the two last classes are shown
only in the right panel.\textbf{ }The red curves show the new empirical
separations defined in the text: between star-forming galaxies and
AGNs (Eq.~\ref{eq:new1}), between Seyfert 2 and LINERs (Eq.~\ref{eq:new3}),
between star-forming galaxies and SF/Sy2 (Eq.~\ref{eq:new2}). The
black dashed curves delimits the region where lies the majority of
composites (SF-LIN/comp region, Eq.~\ref{eq:ineq}).}

\label{fig:newclassif}
\end{figure*}

We define an empirical boundary that allows one to distinguish Seyfert
2 from LINERs in the AGNs region of the blue diagram. It is shown
as a solid diagonal line in Fig.~\ref{fig:newclassif} and follows
the equation below:\begin{equation}
\log([\mathrm{O}\mathsc{iii}]/\mathrm{H}\beta)=0.95\times\log([\mathrm{O}\mathsc{ii}]/\mathrm{H}\beta)-0.4.\label{eq:new3}\end{equation}
This separation minimizes the number of misclassifications between
these two classes. Only 8\% of the Seyfert 2 and 4\% of the LINERs,
according to the red classification, are misclassified respectively
as LINERs or Seyfert 2, according to the new blue classification.

\section{An example application\label{sec:An-example-application}}

\begin{figure*}
\begin{centering}
\includegraphics[width=0.49\textwidth]{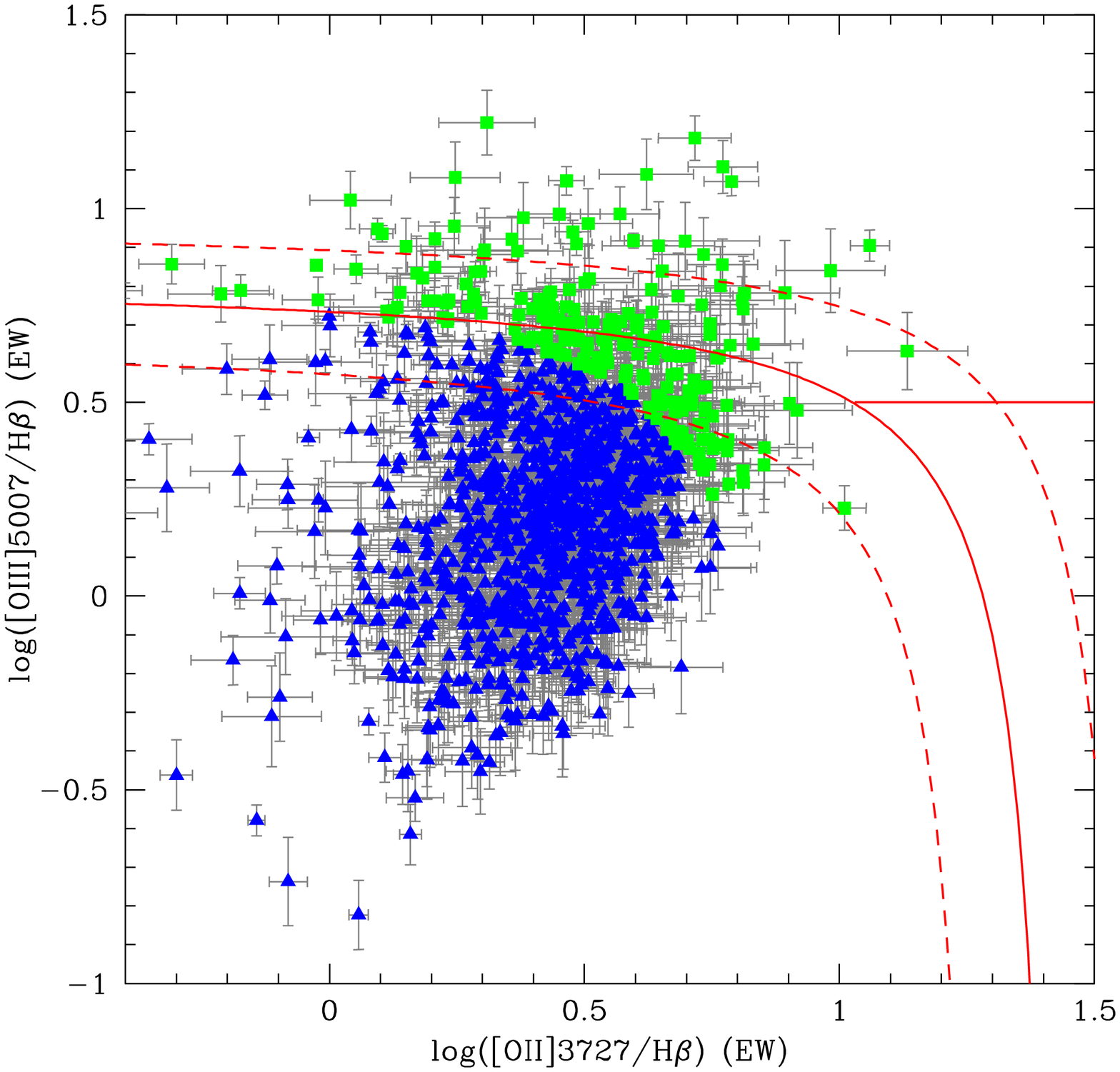} \includegraphics[width=0.49\textwidth]{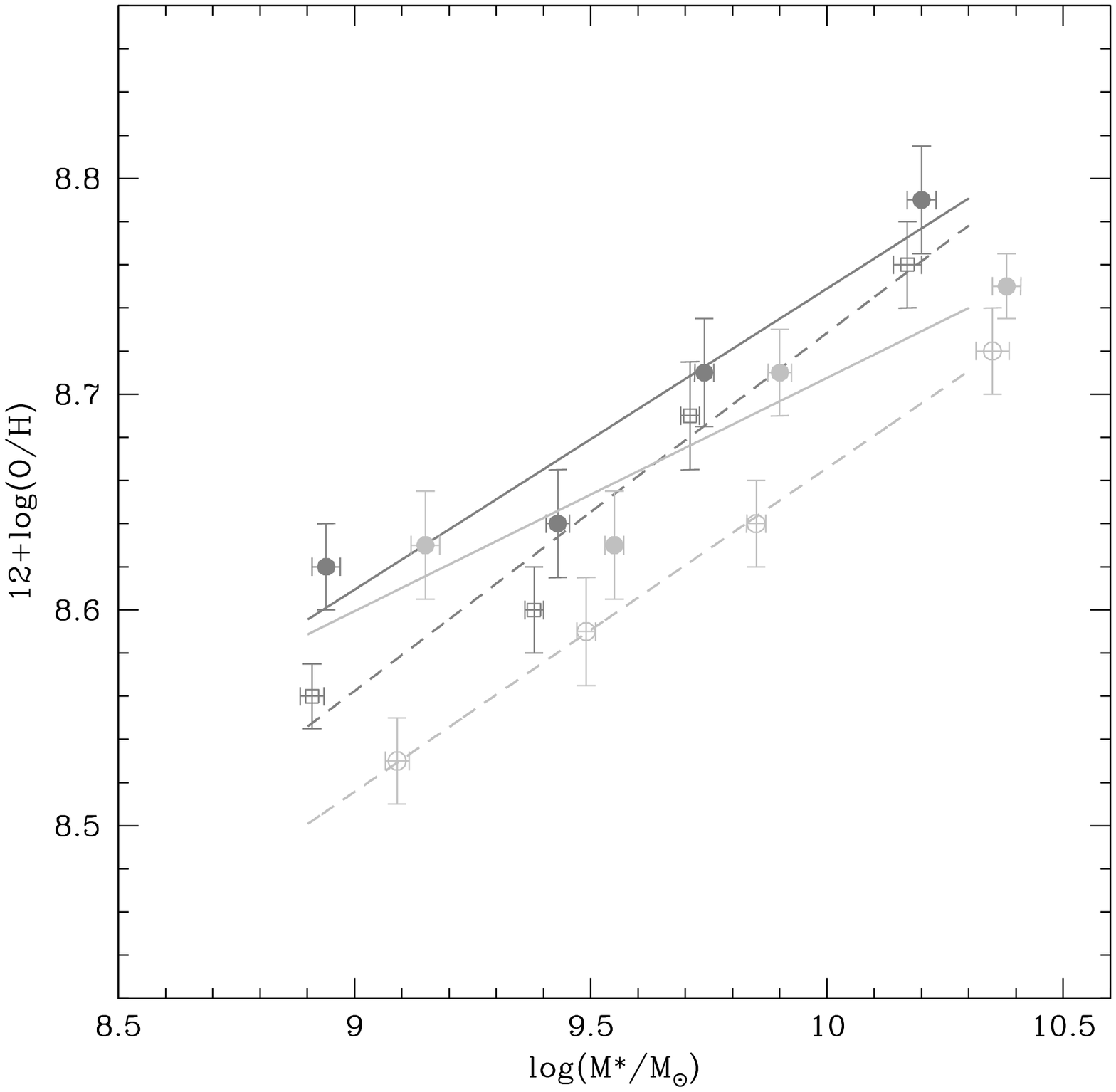} 
\par\end{centering}

\caption{\emph{Left}: Spectral classification of VVDS galaxies \citep{2009A&A...495...53L}.
According to our new classification given in Eq.~\ref{eq:new1},
we classify these objects as star-forming galaxies (blue triangles),
and AGNs (green squares). The red curves shows the older classification
scheme by \citet[see also Fig.~\ref{fig:blueclassif}]{2004MNRAS.350..396L}
\emph{Right}: Relation between the logarithm of the stellar mass of
the gas-phase oxygen abundance of star-forming galaxies in the VVDS
sample. Dashed lines and open symbols show the old results as given
in \citet{2009A&A...495...53L}. Solid lines and filled symbols show
the new results with our new classification. The results are given
in the $0.5<z<0.6$ (dark gray) and $0.6<z<0.8$ (light gray) redshift
ranges.}

\label{fig:mz}
\end{figure*}

Our new classification will be useful for building samples of star-forming
galaxies at high redshift. As an example, we show in Fig.~\ref{fig:mz}
updated results obtained for VVDS data with our new classification.
We refer the reader to \citet{2009A&A...495...53L} for details. The
left panel shows that we now obtain less star-forming galaxies than
with the previous classification scheme. In the right panel, this
ends up to slightly different estimates of the mass-metallicity relation
in the $0.5<z<0.6$ and $0.6<z<0.8$ redshift ranges. With our improved
classification, we now confirm and even strengthen the conclusion
of \citet{2009A&A...495...53L} that the metallicity evolution of
star-forming galaxies is less significant as a function of redshift
for lower mass galaxies than for high mass galaxies.

For galaxies of masses $\sim10^{9}$$M_{\odot}$, the metallicity
evolution is respectively $0.06$dex and $0.10$dex lower in the two
above mentioned redshift ranges than the results obtained with the
old classification of star-forming galaxies. The difference between
the old and the new classification is not significant for galaxies
of masses $\sim10^{10.2}M_{\odot}$.

\section{Conclusion}

\begin{table}
\caption{This table gives the number of galaxies in each class of the reference
classification (columns) as a function of each class of the new blue
classification (lines). Please note that the SF/Sy2 and SF-LIN/comp
are already counted in one of the three main classes (i.e. star-forming
galaxies, Seyfert 2 or LINERs) which may be summed to get the total
number of objects. We use the following abbreviations: SFG: star-forming
galaxies; Sey. 2: Seyfert 2.}

\begin{tabular}{rrrrrr}
\hline 
\hline
 & total & SFG & Sey. 2 & LINER & comp.\tabularnewline
total & 89\,379 & 67\,778 & 2\,949 & 4\,912 & 13\,740\tabularnewline
SFG & 80\,312 & 67\,539 & 952 & 270 & 11\,551\tabularnewline
Sey. 2 & 2\,016 & 55 & 1\,748 & 187 & 26\tabularnewline
LINER & 7\,051 & 184 & 249 & 4\,455 & 2\,163\tabularnewline
SF/Sy2 & 3\,603 & 2\,668 & 774 & 6 & 155\tabularnewline
SF-LIN/comp & 47\,461 & 37\,669 & 17 & 988 & 8\,787\tabularnewline
\hline
\end{tabular}

\label{tab:numbers}
\end{table}

Table~\ref{tab:numbers} summarizes the distribution of the objects
in the new blue classification, as compared to the reference red classification.
A large majority of star-forming galaxies are correctly classified
with the new blue classification. But the new blue classification
also suffer from a non negligible contamination of the star-forming
regions by composites, which should be taken into account in studies
of star-forming galaxies. We note however that a number of composites,
according to the classification of \citet{2003MNRAS.346.1055K}, may
actually only be star-forming galaxies. \citet{2001ApJ...556..121K}
have shown from theoretical modeling that pure star-forming spectra
can be expected in the region latter defined as the composite region
by \citet{2003MNRAS.346.1055K}. \citet{2006MNRAS.371..972S} have
also shown that the composites allow an AGN contribution up to 20\%,
but this does not mean that this contribution cannot be lower than
20\%, or even be zero.\textbf{ }True composites may be confirmed only
from far infrared or X-ray observations. 

We define the region of SF-LIN/comp which contains the majority of
actual composites, but is dominated by actual star-forming galaxies
and LINERs. The composites also contaminates the region of LINERs
in our new classification. However, most of the actual LINERs are
correctly classified with our new classification. 

Finally, the region of Seyfert 2 in our new classification is almost
only composed of actual Seyfert 2 with no significant contamination.
But about a third of the actual Seyfert 2 are\textbf{ }just classified
as SF/Sy2 in our new classification, this class is unfortunately dominated
by actual star-forming galaxies. The DEW classification proposed by
\citet{2006MNRAS.371..972S} actually complements our classification
in classifying correctly SF/Sy2 galaxies. This will be commented in
further details in the second paper of this series. Our new classification
can be used to define samples of star-forming galaxies (see Sect.~\ref{sec:An-example-application}),
but also samples of Seyfert 2 or LINERs (e.g. to compute luminosity
functions) in a much more accurate way than the previous blue classification
scheme. However it cannot be used to derive samples of composites
since they get mixed with star-forming galaxies and LINERs.
\begin{acknowledgements}
The author thanks {}``La Cité de l'Espace'' for financial support
while this paper was being written. The data used in this paper were
produced by a collaboration of researchers (currently or formerly)
from the MPA and the JHU. The team is made up of Stéphane Charlot
(IAP), Guinevere Kauffmann and Simon White (MPA), Tim Heckman (JHU),
Christy Tremonti (Max-Planck for Astronomy, Heidelberg - formerly
JHU) and Jarle Brinchmann ( Sterrewach Leiden - formerly MPA). I thank
in particular J. Brinchmann to keep always these data up to date,
and for his quick answers to my questions. I thank also E. Hache and
J. Marocco, whose work is partly at the origin of the idea to write
this paper. I thank finally the referee G. Stasinska for useful comments
and suggestions, and E. Davoust for improving English. All data presented
in this paper have been processed with the JClassif software, part
of the Galaxie pipeline.

\bibliographystyle{aa}
\bibliography{agn}

\begin{thebibliography}{11}
\expandafter\ifx\csname natexlab\endcsname\relax\def\natexlab#1{#1}\fi

\bibitem[{{Baldwin} {et~al.}(1981){Baldwin}, {Phillips}, \&
  {Terlevich}}]{1981PASP...93....5B}
{Baldwin}, J.~A., {Phillips}, M.~M., \& {Terlevich}, R. 1981, \pasp, 93, 5

\bibitem[{{Heckman}(1980)}]{1980A&A....87..152H}
{Heckman}, T.~M. 1980, \aap, 87, 152

\bibitem[{{Kauffmann} {et~al.}(2003){Kauffmann}, {Heckman}, {Tremonti},
  {Brinchmann}, {Charlot}, {White}, {Ridgway}, {Brinkmann}, {Fukugita}, {Hall},
  {Ivezi{\'c}}, {Richards}, \& {Schneider}}]{2003MNRAS.346.1055K}
{Kauffmann}, G., {Heckman}, T.~M., {Tremonti}, C., {et~al.} 2003, \mnras, 346,
  1055

\bibitem[{{Kewley} {et~al.}(2001){Kewley}, {Dopita}, {Sutherland}, {Heisler},
  \& {Trevena}}]{2001ApJ...556..121K}
{Kewley}, L.~J., {Dopita}, M.~A., {Sutherland}, R.~S., {Heisler}, C.~A., \&
  {Trevena}, J. 2001, \apj, 556, 121

\bibitem[{{Kewley} {et~al.}(2006){Kewley}, {Groves}, {Kauffmann}, \&
  {Heckman}}]{2006MNRAS.372..961K}
{Kewley}, L.~J., {Groves}, B., {Kauffmann}, G., \& {Heckman}, T. 2006, \mnras,
  372, 961

\bibitem[{{Lamareille} {et~al.}(2009){Lamareille}, {Brinchmann}, {Contini},
  {Walcher}, {Charlot}, {P{\'e}rez-Montero}, {Zamorani}, {Pozzetti},
  {Bolzonella}, {Garilli}, {Paltani}, {Bongiorno}, {Le F{\`e}vre}, {Bottini},
  {Le Brun}, {Maccagni}, {Scaramella}, {Scodeggio}, {Tresse}, {Vettolani},
  {Zanichelli}, {Adami}, {Arnouts}, {Bardelli}, {Cappi}, {Ciliegi}, {Foucaud},
  {Franzetti}, {Gavignaud}, {Guzzo}, {Ilbert}, {Iovino}, {McCracken}, {Marano},
  {Marinoni}, {Mazure}, {Meneux}, {Merighi}, {Pell{\`o}}, {Pollo}, {Radovich},
  {Vergani}, {Zucca}, {Romano}, {Grado}, \& {Limatola}}]{2009A&A...495...53L}
{Lamareille}, F., {Brinchmann}, J., {Contini}, T., {et~al.} 2009, \aap, 495, 53

\bibitem[{{Lamareille} {et~al.}(2004){Lamareille}, {Mouhcine}, {Contini},
  {Lewis}, \& {Maddox}}]{2004MNRAS.350..396L}
{Lamareille}, F., {Mouhcine}, M., {Contini}, T., {Lewis}, I., \& {Maddox}, S.
  2004, \mnras, 350, 396

\bibitem[{{Rola} {et~al.}(1997){Rola}, {Terlevich}, \&
  {Terlevich}}]{1997MNRAS.289..419R}
{Rola}, C.~S., {Terlevich}, E., \& {Terlevich}, R.~J. 1997, \mnras, 289, 419

\bibitem[{{Stasi{\'n}ska} {et~al.}(2006){Stasi{\'n}ska}, {Cid Fernandes},
  {Mateus}, {Sodr{\'e}}, \& {Asari}}]{2006MNRAS.371..972S}
{Stasi{\'n}ska}, G., {Cid Fernandes}, R., {Mateus}, A., {Sodr{\'e}}, L., \&
  {Asari}, N.~V. 2006, \mnras, 371, 972

\bibitem[{{Tresse} {et~al.}(1996){Tresse}, {Rola}, {Hammer}, {Stasi{\'n}ska},
  {Le Fevre}, {Lilly}, \& {Crampton}}]{1996MNRAS.281..847T}
{Tresse}, L., {Rola}, C., {Hammer}, F., {et~al.} 1996, \mnras, 281, 847

\bibitem[{{Veilleux} \& {Osterbrock}(1987)}]{1987ApJS...63..295V}
{Veilleux}, S. \& {Osterbrock}, D.~E. 1987, \apjs, 63, 295

\end{thebibliography}

\end{acknowledgements}

\end{document}